\documentclass[letter,twocolumn]{jpsj3}

\usepackage{txfonts}

\usepackage{amsmath,amssymb}
\usepackage{empheq}
\usepackage{graphicx}
\usepackage{bm}
\usepackage{color}

\usepackage{ulem}

\newcommand{\hh}{ \bm{h} }
\newcommand{\kk}{ \bm{k} }
\newcommand{\rr}{ \bm{r} }
\newcommand{\QQ}{ \bm{Q} }

\newcommand{\bdelta}{ \bm{\delta} }

\title{Quadrupole Orders on the fcc Lattice}

\author{Hirokazu Tsunetsugu$^1$, 
Takayuki Ishitobi$^2$ and 
Kazumasa Hattori$^2$}
\inst{$^1$
The Institute for Solid State Physics, 
The University of Tokyo, 
Kashiwanoha 5-1-5, Chiba 277-8581, Japan 
\\
$^2$ 
Department of Physics, 
Tokyo Metropolitan University, 
Hachioji, Tokyo 192-0397, Japan
} 

\abst{
We theoretically study electric quadrupole orders 
in f-electron systems on the fcc lattice.  
Qudrupole degrees of freedom $O_{20}$ and $O_{22}$ 
originate in the non-Kramers doublet ground state $\Gamma_3$ 
of ion with $f^2$ electron configuration.  
For discussing quadrupole orders, we use a minimal model with 
isotropic ($J$) and anisotropic ($K$) 
nearest-neighbor interactions, 
and determine the phase diagram using a four-site 
mean-field approximation at zero and finite temperatures.  
Quadrupoles couple the $\Gamma_3$ doublet to the singlet excited state 
$\Gamma_1$, and its effects on canting antiferro orders 
are examined in detail.  
We found that this coupling leads to a rich phase diagram 
including two- and four-sublattice antiferro phases, 
and that two phases show a partial order of quadrupoles 
at finite temperatures. 
}

\begin{document}
\maketitle

Several years ago, the authors studied antiferro quadrupole orders 
in Pr$\,$1-2-20 compounds\cite{Hattori14,Hattori16,Ishitobi},  
where Pr ions form a diamond sublattice. 
Pr$\,$1-2-20 systems show various exotic phenomena such as 
non Fermi liquids, superconductivity, 
and multipolar phase transitions.\cite{reviewOnimaru}
Each Pr$^{3+}$ ion has two $f$-electrons and 
its ground state is a non-Kramers doublet $\Gamma_3$ 
in cubic environment.  
In $\Gamma_3$ doublet, electric quadrupole has two active components 
O$_{20}$ and O$_{22}$, and they form a two-dimensional 
basis $\QQ$=$(Q_u , Q_v)$ of $E$ irreducible representation 
of the cubic point group. 
They show a long-range order at low temperatures 
into {\it e.g.,} ferro phase for 
PrTi$_2$Al$_{20}$\cite{Sakai,Taniguchi,Kittaka}, 
antiferro phase for PrIr$_2$Zn$_{20}$\cite{Onimaru,Iwasa}, 
and density-wave phase for PrPb$_3$\cite{OnimaruPrPb3,YSatoPrPb3}.  

From a theoretical view point, 
one expects many similarities between these quadrupole systems 
and easy-plane magnets since both groups have a two-component 
order parameter, but two points sharply distinguish between their 
order parameters.  
One is about the time-reversal symmetry: 
$\QQ$ has an even parity while magnetization's is odd. 
The other is their transformation rule 
upon point-group symmetry operations, 
since $\QQ$ is a second-rank tensor while magnetization is an axial vector.  
Therefore, despite quite many theoretical studies on planar 
magnets, new studies are necessary for clarifying characteristics 
of quadrupole orders.\cite{Shiina99} 

Recently, Kusanose \textit{et al.} have studied 
another compound PrMgNi$_4$ 
and discussed an antiferro quadrupole order.\cite{Kusanose19,Kusanose20}  
This material has Pr ions on a fcc sublattice, 
and we will show that 
this difference in lattice structure has important 
implications for quadrupole orders.  
This is due to different effective quadrupole 
interactions between the two lattice structures.  
In the diamond lattice, 
Pr-Pr nearest-neighbor bonds point to [111] and equivalent 
directions, and this limits interactions to 
isotropic ones $J \bm{Q} (\rr )\cdot \bm{Q} (\rr ')$.\cite{Hattori14}  
This isotropic interaction has been used for effective Hamiltonians  
modeling the Pr$\,$1-2-20 system.\cite{Hattori14,Hattori16,%
SBLee,Freyer18,Freyer20}

\begin{figure}[bt]
\begin{center}
\includegraphics[width=8cm]{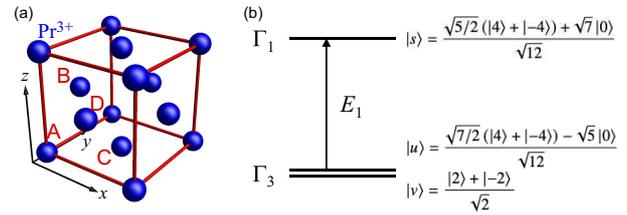}
\end{center}
\caption{
(color online). 
(a) fcc lattice of Pr ions and its four sublattices 
labeled by A-D. 
$\overrightarrow{\mathrm{AB}}$=$\bm{\delta}_1$, 
$\overrightarrow{\mathrm{AC}}$=$\bm{\delta}_5$, 
and 
$\overrightarrow{\mathrm{AD}}$=$\bm{\delta}_3$.  
(b) Non-Kramers ground-state doublet $\Gamma_3$ and 
singlet excited state $\Gamma_1$ of Pr$^{3+}$ ion. 
$| m \rangle$ is the eigenstate $J_z |m \rangle = m |m \rangle$ 
in the $J$=4 multiplet. 
}
\label{fig1}
\end{figure}

We have performed symmetry analysis and found that 
quadrupole interactions on the fcc lattice generally have anisotropic 
terms in addition to isotropic ones. 
Each site is surrounded by twelve nearest neighbors 
separated by 
$\bm{\delta}_{1,2} = (0,1,\pm 1)/2$, 
$\bm{\delta}_{3,4} = (\pm 1,0,1)/2$, 
$\bm{\delta}_{5,6} = (1,\pm 1,0)/2$, 
and their counterparts $-\bm{\delta}$'s 
as shown in Fig.~\ref{fig1}(a).  
The minimal model of quadrupole interactions 
\textit{in $\Gamma_3$ levels} reads as 
\begin{align}
H_Q &= 
\sum_{\rr}
\sum_{j=1}^6
 \QQ (\rr) \cdot \Bigl[
J \mathsf{1} 
+ K \mathsf{g} (\bm{\delta}_j ) \Bigr] \, 
\QQ (\mbox{$\rr$$+$$\bm{\delta}_j$}) 
 ,
\label{eq:ham}
\end{align}
where $K$ is the anisotropic coupling constant 
and $\mathsf{1}$ is the two-dimensional identity matrix.  
Another parametrization 
$(J,K)$=$\bar{J}(\cos \xi , \sin \xi)$ will be also used in this paper.  
Here, the $\rr$-sum is taken over all the sites in the fcc lattice and 
$\QQ$ operates in the $\Gamma_3$ doublet as 
$Q_u$=$ |u \rangle \langle u| - |v \rangle \langle v|$ 
and 
$Q_v$=$ -|u \rangle \langle v| - |v \rangle \langle u|$ 
in terms of the basis states shown in Fig.~\ref{fig1}(b).  
The anisotropy factor is defined as 
\begin{equation}
\mathsf{g} (\bm{\delta} )
= 
 \cos \zeta (\bm{\delta} ) \, \hat{\sigma}_3
-\sin \zeta (\bm{\delta} ) \, \hat{\sigma}_1,
\end{equation}
where 
$\hat{\sigma}_{1,3}$ are Pauli matrices, 
and 
$\bigl( \cos \zeta (\bm{\delta}) , \sin \zeta (\bm{\delta}) \bigr)$
$\propto$ 
$\bigl(
 \mbox{$2$$z^2$$-$$ x^2$$-$$ y^2$}$, 
$\mbox{$\sqrt3$$($$x^2$$-$$y^2$$)$}$ 
$\bigr)$ 
for 
$\bm{\delta}$=$(x,y,z)$.  
Note that 
$\zeta ( \pm \bdelta_{5,6})$=$\pi$, 
and 
$\zeta (\pm \bdelta_{3,4})$=$ - \zeta (\pm \bdelta_{1,2})$=$ \pi/3$.  

Let us first find a classical ground state of this Hamiltonian.  
Performing Fourier transformation, the Hamiltonian is represented as 
$H_Q$=$\sum_{\kk} 
\QQ _{\kk} \cdot 
[ J \gamma_0 (\kk ) 
+ K \bm{\gamma} (\kk ) ] \QQ _{-\kk} 
$
with the coefficients 
$\gamma_0 (\kk )$=%
$2 (\mbox{$c_x c_y$$+$$c_y c_z$$+$$c_z c_x$} )$, 
$\bm{\gamma} (\kk)$=
$\mbox{$\bigl[($$c_x$$+$$c_y$)$c_z$$-$$2 c_x c_y$%
$\bigr]$}$
$\mbox{$\hat{\sigma}_3$$-$%
$\sqrt3$($c_x$$-$$c_y$)$c_z$$\hat{\sigma}_1$}$, 
where $c_a$$\equiv$$\cos (k_a /2)$.  
This effective model was obtained with the special value $K$=$J$
by Kubo and Hotta\cite{Kubo17} 
starting from a microscopic Hamiltonian. 
The special $K$ value is due to a simple form 
of the used microscopic Hamiltonian, but 
various other types of super-exchange processes  
generate $K$$\ne$$J$. 

\begin{figure}[bt]
\begin{center}
\includegraphics[width=8cm]{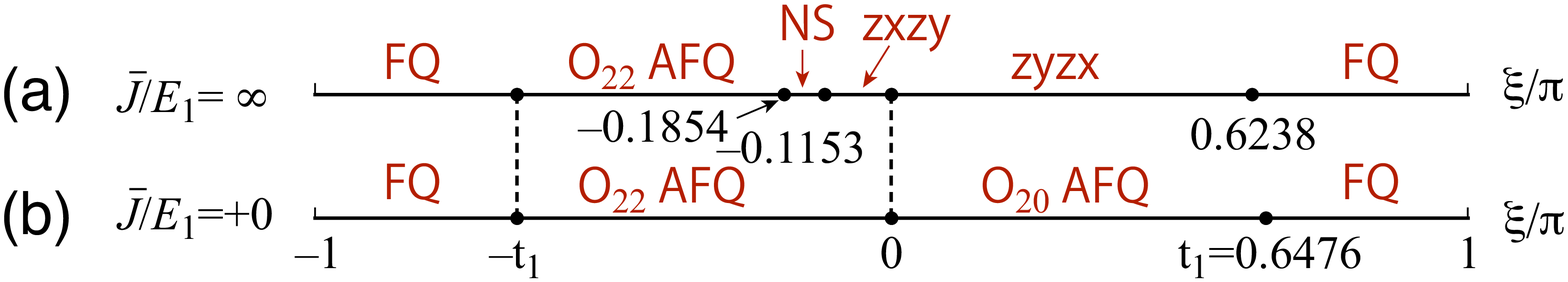}
\\
\vspace{0.3cm}
\includegraphics[width=7cm]{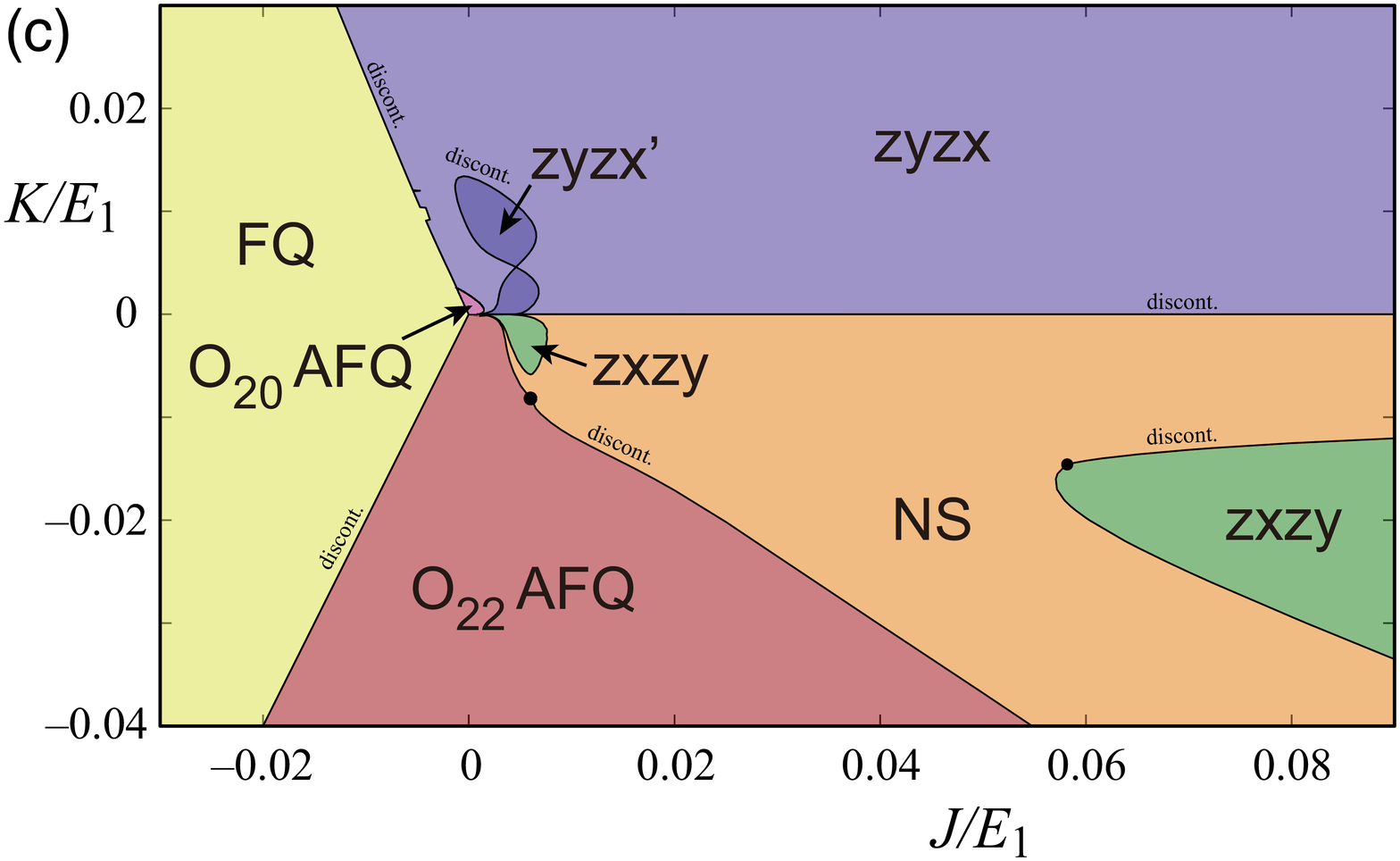}
\end{center}
\caption{
(color online).  
Ground-state phase diagram determined 
by a 4-sublattice mean field theory.
(a) Results in the limit $\bar{J}/E_1$=$\infty$.  
(b) The opposite limit $\bar{J}/E_1$=$+0$, 
and this corresponds to the model $H_Q$.  
$t_1$ is the point where $K$=$-2J$$>$0.  
(c) Results in the $J$-$K$ space. 
Black dots are tri-critical points separating 
continuous and discontinuous transitions. 
$zyzx'$ phase is a special case that $| \langle \QQ \rangle |$ 
is pinned to 1 in one sublattice.  
}
\label{fig2}
\end{figure}

A classical ground state is a spiral state with the wavevector 
$\kk$ where the coefficient matrix $J \gamma_0 + K \bm{\gamma}$ 
has the minimum eigenvalue.  
We can easily show that the ordering wavevector 
is $\kk_X$=(0,0,$2\pi$) when $|K|$$+$$2J$$>$0, 
and $\kk_0$=(0,0,0) when $2J$$<$$|K|$$<$$-$$2J$.  
They correspond to antiferro quadrupole (AFQ) and ferro quadrupole (FQ) 
order, respectively.  
Judging from the corresponding eigenvector, 
the AFQ order parameter is 
$ (-1)^{2z}\, \langle Q_u (\rr )\rangle$ for $K$$>$0 and 
$ (-1)^{2z}\, \langle Q_v (\rr )\rangle$ for $K$$<$0, 
which are A-type antiferro order of O$_{20}$ and O$_{22}$ 
moments, respectively.  
Each AFQ phase has the degeneracy 6=3$\times$2: 
the factor 3 comes from the equivalence of $\kk_X$ with 
$(2\pi,0,0)$ and $(0,2\pi,0)$, while the factor 2 
relates to the exchange of $\pm$ signs.  
Note that the use of a different $\kk$ also requires a mirror 
operation in $(Q_u, Q_v)$ space,\cite{Hattori14} 
since both come from a same $\pi/2$-rotation operation in $\rr$-space.  
In the ferro phase, the $K$-term contribution vanishes and 
the quadrupole internal space is isotropic. 
Therefore, the direction of FQ moment is arbitrary, 
but we will show later that this isotropy is broken by 
the coupling to $\Gamma_1$ excited state. 
We have confirmed that the mean-field theory predicts 
the same ground-state phase diagram, 
and the result is shown in Fig.~\ref{fig2}(b).  
We have also studied 
the related finite temperature phase transition 
using the standard mean field theory.  
The transition temperature is $T_c$=12$|J|$ for FQ phase, 
while $T_c$=4$J$+8$|K|$ for AFQ phase irrespective 
of the sign of $K$. 
Phase transitions at finite temperatures are all second order 
except for bi-critical points $|K| = - 2 J$ and $T_{b} = 6|K|$.

The $K$$>$0 and $K$$<$0 parts of the phase diagram 
are related by a symmetry of the Hamiltonian \eqref{eq:ham} 
at both $T$=0 and $T$$>$0.  
The $\pi/4$-rotation in $\Gamma_3$ Hilbert space transforms 
the operators as 
$\QQ$$\rightarrow$$i\hat{\sigma}_2$$\QQ$.  
Operated by this transformation, the Hamiltonian remains invariant 
except for the sign change $K$$ \rightarrow$$-$$K$, 
since 
$(i\hat{\sigma}_2)^\dagger$$\mathsf{g}(\rr )$%
$(i\hat{\sigma}_2)$=$-$$\mathsf{g}(\rr)$.  
This explains the symmetry in the phase diagram.  
We will see below that coupling to the $\Gamma_1$ excited state 
breaks this symmetry.  

\begin{figure}[bt]
\begin{center}
\includegraphics[width=8cm]{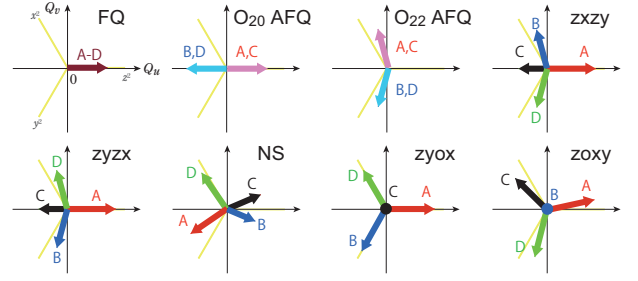}
\end{center}
\caption{
Ordered phases of quadrupoles.  
Typical sublattice $\langle \QQ \rangle$-configurations are shown, 
and color distinguishes different sublattices A-D. 
Yellow lines show Z$_3$ principle axes corresponding to 
the symmetries $3z^2$$-$$r^2$, 
$3x^2$$-$$r^2$, and 
$3y^2$$-$$r^2$.  
Note moments in the $zoxy$ phase are generally inclined from 
the Z$_3$ axes. 
}
\label{fig3}
\end{figure}

Now, let us study the effects of the singlet excited state $\Gamma_1$, 
and denote its energy by $E_1$.  
An important new feature is an induced Z$_3$ anisotropy in 
the internal $\QQ$ space, and this modifies the ordered states.
We have studied this issue for Pr$\,$1-2-20 compounds, for which $K$=0 and 
showed that a parasite FQ moment emerges in AFQ orders.\cite{Hattori14}  
Quadrupole operators connect $\Gamma_3$ doublet with $\Gamma_1$ singlet 
$|s\rangle$, 
and thus the Hamiltonian \eqref{eq:ham} should be modified 
as 
\begin{align}
& H = 
E_1 \sum_{\rr} |s (\rr ) \rangle \langle s (\rr ) | 
+ H_{Q} \bigl[\QQ  \rightarrow 
\QQ ^{(\Gamma_3 + \Gamma_1 )} \bigr] , 
\\[-4pt]
&Q_w ^{(\Gamma_3 + \Gamma_1 )} = 
Q_w 
+  \alpha\, \bigl[
|s \rangle \langle w | + \mbox{h.c.} \bigr] , \ 
\ \ \alpha \equiv {\textstyle \frac{\sqrt{35}}{2}} ,
\end{align}
for $w$=$u$,$v$.\cite{Hattori14} 
In the following, the quadrupole operators $\QQ$ refer to 
those generalized $\QQ ^{(\Gamma_3 + \Gamma_1 )}$.  
It is clear that the new operators do not have the aforementioned 
symmetry upon $\pi/4$-rotation in $\Gamma_3$ space.  

As before, let us examine symmetry breaking in the ground state 
by the mean field theory.  
A crucial difference from the previous calculation is that now 
the response $\langle \QQ \rangle$ 
generally does not parallel its molecular field 
$\hh$$=:$$h \, (\cos \eta , \sin \eta )$.  
When $h$$\ll$$E_1$, one can use the degenerate perturbation theory 
and calculate the response. 
Consider a single-site mean field Hamiltonian 
$H_\mathrm{mf}$$=$$E_1 |s \rangle \langle s |$$-$$\hh \cdot \QQ$. 
Its ground state is 
$|\psi_0 \rangle \approx 
\cos (\eta /2) |u \rangle 
- \sin (\eta /2) |v \rangle 
+ (\alpha h / E_1 ) \cos (3 \eta /2) |s \rangle
$, 
and the response is immediately obtained
\begin{align}
\left[ \begin{array}{c}
\!\! \langle Q_u \rangle \!\!
\\
\!\! \langle Q_v \rangle \!\!
\end{array}
\right] 
\approx 
\frac{\hh}{h} + \chi_1 \hh 
+ \chi_1 h 
\left[ \begin{array}{c}
\!\! \phantom{-} \cos 2\eta \!\! 
 \\
\!\! -\sin 2\eta \!\! 
\end{array} \right] , 
\ \ \ 
\chi_1 \equiv \frac{\alpha^2}{E_1} . 
\end{align}
The last term is not parallel to $\hh$, unless 
$\eta$=$(\pi /3) \times (\mbox{integer})$.  
This manifests that the $\QQ$-space symmetry is reduced 
down to Z$_3$, 
compared with the O(2) symmetry in the mean field approximation 
of $H_Q$.  

In the two-sublattice AFQ phase, the molecular fields of the A- and 
B-sublattices 
are related to the two order parameters as 
$\hh ^A$=$
- 4 ( J - K \hat{\sigma}_3 ) 
\langle \QQ ^A \rangle$$- 4 (2J + K \hat{\sigma}_3 ) 
\langle \QQ ^B \rangle$, 
and $\QQ ^A$ and $\QQ ^B$ are exchanged for $\hh ^B$.  
The O$_{20}$ AFQ phase has the solution 
$\langle \QQ ^A \rangle \approx (1 + 35 (J+2K)/E_1 , 0)$ 
with $\langle Q_u^B \rangle$=$(-1,0)$ unchanged.  
The FQ phase's solution is 
$\langle \QQ \rangle \approx (1 + 210 |J|/E_1 ,0)$. 
Note that the order parameter angle points to one of 
the Z$_3$ axes, where $|\langle \QQ \rangle|/|\hh |$ is maximum.  
Calculation for the O$_{22}$ AFQ phase is more elaborate.  
Since $\hh ^{A,B}$ and $\langle \QQ^{A,B} \rangle$ are not parallel, 
one needs to determine 
the tilting of the molecular field 
$\eta^{A,B}$=$\pm (\pi /2 + \delta \eta )$ 
self-consistently. 
The result is 
\begin{align}
&\langle \QQ ^A \rangle \approx 
\left[
\begin{array}{c}
\!\!\!\!\!\!\!\!\!\! \delta \eta  \!\!\!\!\!\!\!\!\!\! 
\\ 
1 
\end{array}
\right]
+ 35 \frac{\mbox{$J$+$2|K|$}}{E_1} 
\left[
\begin{array}{c}
\!\!\! -1 \!\!\! \\ \!\! 1 \!\! 
\end{array}
\right] , 
\ \ 
\delta \eta 
= 
\frac{105J(\mbox{$J$+$2|K|$})}{2(\mbox{$2J$+$|K|$}) E_1} , 
\end{align}
and $\langle \QQ ^B \rangle = \hat{\sigma}_3 \langle \QQ ^A \rangle $. 
Note that $\delta \eta$ diverges with approaching the boundary 
to the FQ phase.  

When the $\Gamma_1$ level is not high, the perturbation theory 
breaks down, and we need numerical calculations 
to solve the mean field equations self-consistently.  
In the FQ phase ($2J$$<$$K$$<$$2|J|$), 
the order parameter $ \langle Q_u \rangle$ 
approaches continuously to the maximum limit $7/2$  
as $ |J|/E_1 \rightarrow \infty$.  
The coupling to $\Gamma_1$ level changes the AFQ orders drastically.  
It destabilizes the 2-sublattice structure such as simple 
O$_{20}$ or O$_{22}$ order, and 4-sublattice orders appear 
in a wide parameter range.  

Let us first examine the limit of strong interactions 
$\bar{J}/E_1$$\rightarrow$$\infty$. 
We have performed a mean-field analysis using four sublattices and 
found 5 phases as shown in Fig.\ref{fig2} (a).  
Recall 
$(J,K)$=$\bar{J}$$(\cos \xi,\sin \xi)$.  
Compared to the weak interaction limit $\bar{J}/E_1$=$+0$, 
the O$_{20}$ phase is mainly replaced by the $zyzx$ phase, 
and a considerable part of the O$_{22}$ phase is replaced 
by the $zxzy$ phase and the nonsymmetric (NS) 4-sublattice phase.  
Sublattice quadrupole structure in these phases is 
illustrated in Fig.~\ref{fig3}.  
These new 4-sublattice orders $zyzx$ and $zxzy$ 
also have a finite degeneracy related to the cubic point group, 
and the degeneracy is 12.  
Low symmetry of the NS phase questions  
its stability in better approximations with a larger 
unit cell, but we leave this issue for a future study.  

The full phase diagram is calculated in $J$-$K$ parameter space 
and shown in Fig.~\ref{fig2}(c).  
It is quite surprising that the O$_{20}$ AFQ phase is 
very narrow and destabilized into the $zyzx$ phase, where 
not only O$_{20}$ but also O$_{22}$ components are nonzero.  
This comes from a special constraint of the $\Gamma_3$+$\Gamma_1$ 
Hilbert space.  
Applied by the molecular field $\bm{q}$=$(q_u,0)$, 
the response $\langle Q_u \rangle$ grows from $+1$ linearly 
if $q_u$$>$0, 
but stays $-1$ if $q_u$$<$0.  
Thus, the O$_{20}$ AFQ phase cannot gain an enough interaction 
energy, and this is the origin of instability.

\begin{figure}[t!]
\begin{center}
\includegraphics[width=0.5\textwidth]{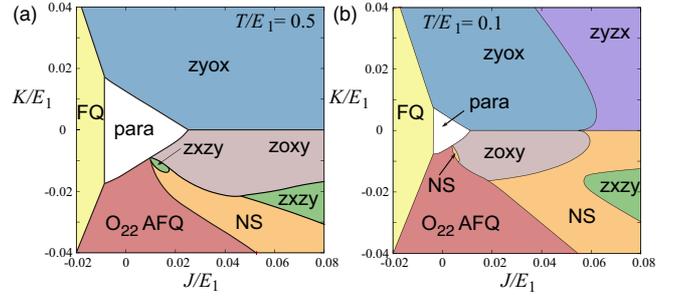}
\end{center}
\vspace{-0.5cm}
\caption{(color online). 
$J$-$K$ phase diagram for 
$T/E_1$$=$(a) 0.5 and (b) 0.1.}
\label{fig:finiteTphase}
\end{figure}

\begin{figure}[t!]
\begin{center}
\includegraphics[width=0.5\textwidth]{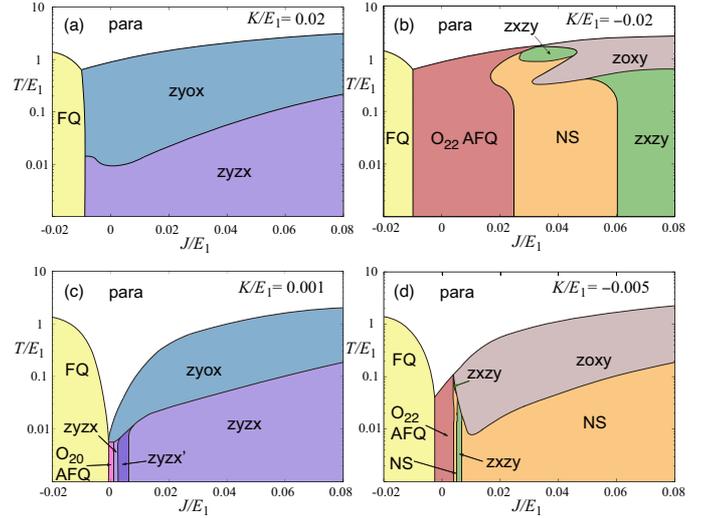}
\end{center}
\vspace{-0.5cm}
\caption{(color online). 
$J$-$T$ phase diagram for 
$K/E_1$$=$(a) 0.02, (b) $-0.02$, (c) $0.001$, and (d) $-0.005$.
``para" denotes the high temperature disordered phase.}
\label{fig:finiteTphase2}
\end{figure}

We now discuss finite temperature properties of quadrupole orders.
At finite temperatures, there appear two new phases which are absent 
in the ground state: the $zyox$ phase 
for $K$$>$0 and the $zoxy$ phase for $K$$<$0 
as shown in Fig.~\ref{fig:finiteTphase}(a) for $T$=$0.5$. 
The two sublattice structures are shown in Fig.~\ref{fig3}, and 
interestingly 
they are a partial-ordered phase in which one sublattice 
remains disordered,
i.e., the quadrupole moment vanishes. 
These two phases emerge through first-order transition 
as $T$ decreases in a wide region of $J$-$K$ parameter space. 
See also Fig.~\ref{fig:finiteTphase2}, 
where the $J$-$T$ phase diagrams for fixed $K$ are shown.  
We will explain later that they are 
triple-${\bm q}$ orders of quadrupoles in the fcc lattice.  

The other regions in the $J$-$K$ plane are occupied by
the phases discussed for $T$=0; the FQ phase for $J$$<$0, 
the O$_{22}$ AFQ order for $K$$<$0 and smaller $J$, 
and the $zxzy$ and the NS phases for larger $J$ and $K$$<$0.
The $zyzx$ phase does not 
appear at $T$=0.5 in the range shown in Fig.~\ref{fig:finiteTphase}(a), 
but gradually starts to dominate at low temperatures 
a large region of the $K$$>$0 part 
as shown in Fig.~\ref{fig:finiteTphase}(b) 
and also in Figs. \ref{fig:finiteTphase2}(a) and (c). 
As shown in Figs.~\ref{fig2}(c) and \ref{fig:finiteTphase2}(c), 
the O$_{20}$ AFQ phase is limited to a tiny region. 
This clearly contrasts with the O$_{22}$ AFQ order for $K$$<$0. 
In addition, the O$_{20}$ AFQ phase does not touch the disordered 
phase (denoted as ``para'' in Figs.~\ref{fig:finiteTphase} 
and \ref{fig:finiteTphase2}), 
while the O$_{22}$ AFQ phase does. 
Indeed, for $K$$>$0, 
a first-order transition generally takes place to 
the partial-ordered phase $zyox$ 
at a temperature higher than that the AFQ order sets in.  
We demonstrate this below.  


Since the uniform component of quadrupole moments vanishes 
in the two triple-${\bm q}$ phase, 
their analysis is easier than the others.  
We can write down the Landau free energy in terms 
of order parameters 
$\{ {\bm Q}_{i} \}_{i=1}^3$. 
They are 
$\langle \QQ ( \mbox{$\bm{k}$=$\bm{q}_i$} )\rangle$ 
defined at the three zone-boundary wavevectors 
${\bm q}_1$=$(2\pi,0,0)$, 
${\bm q}_2$=$(0,2\pi,0)$, and 
${\bm q}_3$=$(0,0,2\pi)$=${\bm k}_X$, and 
also parameterized as 
${\bm Q}_i$=$(Q_{iu},Q_{iv})$%
=$Q_i (\cos\theta_i,\sin\theta_i)$ with $Q_i\ge 0$.
At these $\bm{q}_i$'s,
the exchange interactions 
$ J \gamma_0 ({\bm k})$+$ K \bm{\gamma} ({\bm k} )$ 
have the lowest eigenvalue $-2J-4|K|$ 
and their eigenvectors correspond to  
$(\theta_1,\theta_2,\theta_3)$=$(2\pi/3,4\pi/3,0)$ for $K$$>$0 and 
$(7\pi/6,11\pi/6,\pi/2)$ for $K$$<$0.
Note that the mode for $\theta_i+\pi$ is equivalent to $\theta_i$, 
as far as the harmonic parts are concerned. 
The phase with only one $\QQ_i$$\ne$$\bm{0}$ is a 
natural choice for discussing 
the second-order transition of 
the single-${\bm q}$ order with the highest transition temperature.
One can write down the free energy density 
for general cases as 
\begin{align}
F =& \sum_{i=1}^3
\left({\textstyle \frac12} a Q_i^2 + c Q_i^4 \right) 
+4c' \sum_{i < j}
 Q_i^2 Q_j^2 \cos^2( \mbox{$\theta_i$$-$$\theta_j$})
\nonumber\\[-6pt]
& 
-b Q_1 Q_2 Q_3 \cos\bar{\theta}
+\cdots , 
\hspace{0.7cm}
\left( 
\mbox{$\bar{\theta}$$\equiv$$\theta_1$+$\theta_2$+$\theta_3$}
\right) . 
\end{align}
Since
${\bm q}_1$+${\bm q}_2$+${\bm q}_3$ is a reciprocal lattice vector, 
the third-order term is nonvanishing and plays a crucial role 
in stabilizing the triple-${\bm q}$ phases. 
Note that one can take $b>0$ without loss of generality.

For $K$$>$0, the triple-$\bm{q}$ phase with the choice 
$(\theta_1,\theta_2,\theta_3)$
=$(2\pi/3,4\pi/3,0)$ 
satisfies $\bar{\theta}$=0, and thus 
this gains an energy from the third-order term. 
This demonstrates that the triple-$\bm q$ order takes place 
through a first-order transition with 
a higher transition temperature than the AFQ 
order's where 
$a$=0. 
This is a very unique property of the nonmagnetic 
triple-$\bm{q}$ order with $\bm{q}$'s at the 
Brillouin zone boundary.
Among various possibilities of the triple-${\bm q}$ orders, 
promising candidates are those consisting of the eigen modes 
with the lowest eigenvalue of 
$ J \gamma_0 ({\bm k})$+$ K \bm{\gamma} ({\bm k} )$. 
There are two such candidates.  
One is the triple-${\bm q}$ phase with 
$Q_1$=$Q_2$=$Q_3$ and 
$F$=$a Q^2/2$%
$-$$ 2b Q^3/\sqrt{3} $%
$+$$(c$$+$$c'/3)Q^4$, 
where 
$Q^2$$\equiv$$Q_1^2$$+$$Q_2^2$$+$$Q_3^2$. 
The other is that with $Q_1$=$Q_2$$\ne$$Q_3$=$3b/c'$ and 
$F$=$a Q^2/2$$+$$c Q^4$$+$$c'Q_1^4$. 
While the free energy for O$_{20}$ AFQ is given by 
$F$=$a Q^2/2$$+$$cQ^4$, 
one should not compare this directly to those above 
for the triple-$\bm q$'s. 
This is because the optimal value of $Q_i$'s differs for each phase 
(the triple-${\bm q}$ phases are realized even for 
$a$$>$0). 

Examining the real-space profile 
$Q({\bm r})$$\propto$%
$ \Re \sum_{i} e^{i{\bm q}_i\cdot {\bm r}} \bm{Q}_i$, 
it turns out that the former corresponds to 
the $zyox$ phase 
which is realized when the moment size is small, 
while the latter to 
the $zyzx$ phase
when the moment is larger. 
In Figs.~\ref{fig2}(c) and \ref{fig:finiteTphase2}(c), 
a first-order transition occurs inside the $zyzx$ phase 
\textit{with no symmetry change}, 
and we have named the inner-side phase $zyzx'$. 
The difference lies in the magnitudes of $Q_i$.  
At $T$=0, $Q_{u}^C$ is pinned at $-1$ in the $zyzx'$ phase, 
while positive in the nearby region of the $zyzx$ phase. 
At finite temperatures, 
$Q_1$$=$$Q_2$$>$$Q_3$ in $zyzx'$, 
while $Q_1$$=$$Q_2$$<$$Q_3$ for $zyzx$.
Note that in $zyzx$ and $zyzx'$ phases, the uniform quadrupole moment 
${\bm Q}_0$=$\langle \bm{Q}(\mbox{$\kk$=$\bm{0}$})\rangle$ 
is also finite, 
since another third-order term
$-b' \sum_{i=1}^3 \bigl[ Q_{0u} ( \mbox{$Q_{iu}^2$$-$$Q^2_{iv}$} )$%
$-$$ 2Q_{0v} Q_{iu} Q_{iv}\bigr]$
gains an energy, 
while it vanishes in the $zyox$ phase. 
The direction of ${\bm Q}_0$ is parallel (antiparallel)
to ${\bm Q}_3$ in the $zyzx$ ($zyzx'$) phase.
When integrating out the uniform component ${\bm Q}_0$, 
the form of the free energy is unchanged, 
but the coefficients $c$ and $c'$ are renormalized.

The part of $K$$<$0 requires a more elaborate analysis, 
and we discuss the competition 
between the O$_{22}$ AFQ and $zoxy$ phases 
only qualitatively.
In contrast to the case for the $K$$>$0 part, 
the triple-${\bm q}$ phases consisting of the lowest eigenmode 
for each of ${\bm q}_i$ have no third-order free energy gain, 
since $-b Q_1 Q_2 Q_3 \cos\bar{\theta}$=0 
when substituting 
$(\theta_1,\theta_2,\theta_3)$=$(\pi/2,7\pi/6,11\pi/6)$ 
for $K$$<$0.
Thus, in order to achieve an gain of the third-order term, 
each $\QQ_i$ should be optimized by hybridizing the two eigen modes.  
We should note that incommensurate orders cannot be ruled out here, 
but we leave this analysis for a future study.  

The simplest triple-${\bm q}$ ansatz is 
a symmetric hybridization  
with the same amplitude and ``tilting'' angle 
$\phi$
for all the ${\bm q}_i$:
$\bm{Q}_i$=$%
\bar{Q}  \bigl( 
\cos(\mbox{$\theta_i$$-$$\phi$})$,%
$\sin(\mbox{$\theta_i$$-$$\phi$}) \bigr)$ 
with $\theta_i$=$(\mbox{$4i$$-$1} )\pi/6$.
This choice corresponds to the $zoxy$ phase appearing 
in Fig.~\ref{fig:finiteTphase} for $K$$<$0.   
We find its free energy as 
$F$=$[ \mbox{$a_0 /2$+$4K \cos (2\phi)$}]$%
$\bar{Q}^2$+$b \sin (3\phi) \bar{Q}^3$+$\cdots$ 
by approximating $a(K)$ around $K$=0.  
In contrast, the free energy for O$_{22}$ AFQ with 
${\bm k}$$=$${\bm k}_X$ is 
$( \mbox{$a_0 /2$+${4}K$})$$Q_3^2$+$\cdots$.
Thus, the stability of the $zoxy$ phase is 
controlled by the cost $4K\cos(2\phi)\bar{Q}^2$ 
and the gain $b\sin(3\phi)\bar{Q}^3$. 
For larger $|K|$, as the energy cost increases in the second-order term,  
the $zoxy$ phase is not favored, 
which is consistent with 
the mean-field results for the microscopic model 
as shown in Fig.~\ref{fig:finiteTphase}.

A crucial stabilization mechanism of 
the triple-${\bm q}$ phases such as $zyox$, $zyzx$, and $zoxy$ 
discussed above is the third-order invariant in the free energy. 
This differs strikingly from magnetic systems,  
where it is prohibited by the time-reversal symmetry.
Within the Landau theory, 
the mechanism stabilizing the other phases depends on 
the details of the higher-order terms and the coupling 
with the uniform quadrupole moment ${\bm Q}_0$ 
in the third- and fourth-order terms.

In this paper, we have theoretically studied 
electric quadrupole orders of non-Kramers local doublets $\Gamma_3$ 
states on the fcc lattice.  
To this end, we constructed a minimal effective 
model by taking account of the lattice structure, 
and analyzed it by a four-site mean-field theory. 
Its Hamiltonian has three energy scales: 
isotropic and anisotropic exchange constants 
$J$, $K$, 
and the excitation energy of the $\Gamma_1$ state 
$E_1$. 
It has been known that the ratio 
$\sqrt{J^2 \! + \! K^2}/E_1$ 
measures the strength of the Z$_3$ anisotropy characteristic 
to systems with time-reversal symmetry such as electric quadrupoles.  
The limit of negligible Z$_3$ anisotropy 
($\sqrt{J^2 \! + \! K^2}/E_1$=0) 
has three ordered phases: ferro, and 
two 2-sublattice antiferro phases 
(O$_{20}$- and O$_{22}$-antiferro). 

When the excitation energy $E_1$ is not so large, 
the Z$_3$ anisotropy produces 
considerable effects and stabilizes phases with more 
complex quadrupole structure.  
Two of them are the $zyox$ and $zoxy$ phases, and they have 
4-sublattice antiferro partial orders that quadrupole 
moment is zero in one sublattice.  
These two are also special in that the ferro 
component of quadrupoles vanishes, while the other 
2- and 4-sublattice antiferro phases have a nonzero 
parasitic ferro component.  
One can understand that these partial ordered phases are 
a triple-$\bm{q}$ order that the Z$_3$ anisotropy 
couples the three modes with wavevectors 
$(2\pi , 0, 0)$, $(0, 2\pi , 0)$, and $(0, 0, 2\pi)$. 
The results of our calculations propose quite many 
candidates for the discussed quadrupole antiferro 
order that takes place in the material PrMgNi$_4$.  
A first screening would be the checking whether the cubic lattice 
symmetry is broken or not, and its result narrows the order 
identification.  
The direction of the order parameter $\langle {\QQ } (\rr) \rangle$ 
determines the lowered local symmetry at each Pr ion, and 
probing atomic displacements also helps in identifying the order.

\begin{acknowledgment}
This work was supported by a Grant-in-Aid for Scientific 
Research (Grant Nos.~16H04017 and 18K03522) from the Japan 
Society for the Promotion of Science.

\end{acknowledgment}


\begin{thebibliography}{9}
%
\bibitem{Hattori14}
K.~Hattori and H.~Tsunetsugu, 
J. Phys. Soc. Jpn. \textbf{83}, 034709 (2014). 
%
%
\bibitem{Hattori16}
K.~Hattori and H.~Tsunetsugu, 
J. Phys. Soc. Jpn. \textbf{85}, 094001 (2016). 
%
%
\bibitem{Ishitobi}
T.~Ishitobi and K.~Hattori, 
J. Phys. Soc. Jpn. \textbf{88}, 063708 (2019).
%
%
\bibitem{reviewOnimaru}
T.~Onimaru and H.~Kusunose, 
J. Phys. Soc. Jpn. \textbf{85}, 082002 (2016).
%
%
\bibitem{Sakai} 
A.~Sakai, K.~Kuga, and S.~Nakatsuji, 
J. Phys. Soc. Jpn. \textbf{81}, 083702 (2012).
%
%
\bibitem{Taniguchi} 
T.~Taniguchi, K.~Hattori, M.~Yoshida, H.~Takeda, 
S.~Nakamura, T.~Sakakibara, M.~ Tsujimoto, A.~Sakai, 
Y.~Matsumoto, S.~Nakatsuji, and M.~Takigawa, 
J. Phys. Soc. Jpn. \textbf{88}, 084707 (2019).
%
%
\bibitem{Kittaka} 
S.~Kittaka, T.~Taniguchi, K.~Hattori, S.~Nakamura, T.~Sakakibara, 
M.~Takigawa, M.~Tsujimoto, A.~Sakai, Y.~Matsumoto, and S.~Nakatsuji, 
J. Phys. Soc. Jpn. \textbf{89}, 043701 (2020).
%
%
\bibitem{Onimaru} 
T.~Onimaru, K.~T.~Matsumoto, Y.~F.~Inoue, K.~Umeo, 
T.~Sakakibara, Y.~Karaki, M.~Kubota, and T.~Takabatake, 
Phys. Rev. Lett. \textbf{106}, 177001 (2011).
%
%
\bibitem{Iwasa} 
K.~Iwasa, K.~T.~Matsumoto, T.~Onimaru, T.~Takabatake, 
J-. M.~Mignot, and A.~Gukasov, 
Phys. Rev. B \textbf{95}, 155106 (2017).
%
%
\bibitem{OnimaruPrPb3} 
T.~Onimaru, T.~Sakakibara, N.~Aso, H.~Yoshizawa, 
H.~S.~Suzuki, and T.~Takeuchi
Phys. Rev. Lett. \textbf{94}, 197201 (2005).
%
%
\bibitem{YSatoPrPb3} 
Y.~Sato, H.~Morodomi, K.~Ienaga, Y.~Inagaki, 
T.~Kawae, H.~S.~Suzuki, and T.~Onimaru,  
J. Phys. Soc. Jpn. \textbf{79}, 093708 (2010).
%
\bibitem{Shiina99}
R.~Shiina, H.~Shiba, and O.~Sakai, 
J. Phys. Soc. Jpn. \textbf{68}, 2105 (1999).  
%
%
\bibitem{Kusanose19}
Y.~Kusanose, T.~Onimaru, G.-B.~Park, Y.~Yamane, 
K.~Umeo, T.~Takabatake, N.~Kawata, and T.~Mizuta, 
J. Phys. Soc. Jpn. \textbf{88}, 083703 (2019). 
%
%
\bibitem{Kusanose20}
Y.~Kusanose, T.~Onimaru, G.-B.~Park, Y.~Yamane, 
K.~Umeo, and T.~Takabatake, 
JPS Conf. Proc. \textbf{30}, 011160 (2020). 
%
%
\bibitem{SBLee}
S.~B.~Lee, S.~Trebst, Y.~B.~Kim, and A.~Paramekanti, 
Phys.~Rev.~B \textbf{98}, 134447 (2018). 
%
%
\bibitem{Freyer18}
F.~Freyer, J.~Attig, S.~B.~Lee, A.~Paramekanti, S.~Trebst, and Y.~B.~Kim, 
Phys.~Rev.~B \textbf{97}, 115111 (2018). 
%
%
\bibitem{Freyer20}
F.~Freyer, S.~B.~Lee, Y.~B.~Kim, S.~Trebst, and A.~Paramekanti, 
Phys.~Rev.~Reserach \textbf{2}, 033176 (2020). 
%
%
S.~Nakatsuji, and Y.~B.~Kim, 
Nature Comm. \textbf{10}, 4092 (2019).  
%
%
%
\bibitem{Kubo17}
K.~Kubo and T.~Hotta, 
Phys. Rev B \textbf{95}, 054425 (2017). 
%
%
\end{thebibliography}
\end{document}